\documentclass[12pt]{article}       
\usepackage{graphicx}
\usepackage{amsmath}
\usepackage{amssymb}
\usepackage{amsfonts}
\usepackage{braket}
\usepackage{subfigure}

\usepackage[a4paper,top=2cm,bottom=4cm,left=3.5cm,right=3cm]{geometry} 
\usepackage{fancyhdr}

\title{Cross-diffusion driven instability in a predator-prey system
with cross-diffusion}
\author{E.Tulumello\footnote{\tt{tulunora@gmail.com}}, M.C.Lombardo\footnote{\tt{mariacarmela.lombardo@math.unipa.it}}, M.Sammartino\footnote{\tt{marco.sammartino@math.unipa.it}}\\
 University of Palermo, Department of Mathematics,\\ Via Archirafi 34, 90123 Palermo, Italy.\\
	}

\date{\today}

\begin{document}

\maketitle

\begin{abstract}
In this work we investigate the process of pattern formation induced by nonlinear diffusion in a reaction-diffusion system with Lotka-Volterra predator-prey kinetics. We show that the cross-diffusion term is responsible of the destabilizing mechanism  that leads to the emergence of spatial patterns. Near marginal stability we perform a weakly nonlinear analysis to predict the amplitude and the form of the pattern, deriving the Stuart-Landau amplitude equations. Moreover, in a large portion of the subcritical zone, numerical simulations show the emergence of oscillating patterns, which cannot be predicted by the weakly nonlinear analysis. Finally, when the pattern invades the domain as a travelling wavefront, we derive the Ginzburg-Landau amplitude equation which is able to describe the shape and the speed of the wave.

  
\end{abstract}

\section{Introduction}
The aim of this paper is to investigate the process of pattern formation for the following reaction-diffusion system:
\label{intro}
\begin{equation}
\label{model} 
\begin{split}
\partial_\tau U &=\Gamma U\biggl(R-\frac{R}{K}U-\gamma_{12}V\biggl)+D_1 \Delta U,\\
\partial_\tau V &=\Gamma V(-M+\gamma_{21}U)+D_{21}\nabla \cdot \left (V\nabla U\right)+D_2 \Delta V.
\end{split}
\end{equation}
Here $U(\boldsymbol{z},\tau)$ and $V(\boldsymbol{z},\tau)$, with $\boldsymbol{z}\in \Omega  \subseteq \mathbb{R}^n$, are the population densities of preys and predators, respectively. The kinetics is of the Lotka-Volterra predator-prey type with a logistic term for the preys: $R$ and $K$ are the growth rate and carrying capacity for the prey, respectively, $M$ is the  predator death rate, $\gamma_{12}$ and $\gamma_{21}$ are the rate of predation and the capture efficiency and  $\Gamma$ gives the size of the spatial domain.
The spatial motion of the two populations is described by the usual linear diffusion terms and by a nonlinear term deriving from the assumption that predators avoid zones of high prey density.

Without loss of generality, we shall consider the following dimensionless form of the system \eqref{model} for $u$ and $v$:
\begin{equation}
\label{pp_adim}
\begin{split}
\partial_t u &=\Gamma u(r-\gamma u-v)+\nabla^2u,\\
\partial_t v &=\Gamma v(-1+u)+d_{21}\nabla \cdot(v\nabla u)+d_2 \nabla^2 v,
\end{split}
\end{equation}
where
\begin{equation}
u(\boldsymbol{x},t)=\frac{\gamma_{21}}{M}U,\quad v(\boldsymbol{x},t)=\frac{\gamma_{12}}{M}V\quad\text{with}\quad
t=M\tau, \quad \boldsymbol{x}=\sqrt{\frac{M}{D_1}}\boldsymbol{z},\quad
\end{equation}
and
\begin{equation}
r=\frac{R}{M},\quad  \gamma=\frac{R}{K\gamma_{21}},\quad d_{21}=\frac{D_{21}M}{D_1\gamma_{21}},\quad d=\frac{D_2}{D_1}.
\end{equation}
In what follows we shall restrict ourselves to the 1D domain $\Omega=[0, 2\pi]$ on which we shall impose homogeneous Neumann boundary conditions.



System \eqref{pp_adim} belongs to a large class of models involving density depending diffusion (see e.g. \cite{MRW11}) or cross-diffusion to reproduce segregation effects and the creation of spatial niches \cite{K-S,SKT,Cina3}.
 Cross-diffusion terms should be introduced when the gradient of the density of one species induces a flux of another species. It has been shown that, for a large class of predator-prey or competitive kinetics without an autocalitic term, classical diffusion is not sufficient for  pattern formation, no matter what diffusion rates are: in these cases, cross diffusion is necessary for pattern formation \cite{GLS,GLS2d,Cina3}. It is worth mentioning that the introduction of the cross diffusion terms, other than being simply guessed a priori, can  rigorously be obtained through a self-consistent derivation anchored to the microscopic
world (see \cite{FCD}) or deduced from mutation and splitting of a single species (see \cite{G,CD}). Cross-diffusion terms of the type introduced  in \eqref{pp_adim} have already appeared in \cite{Dubey,Kuto,Cina3} (modeling  prey congregation to protect from the attack of the predator), but also in other contexts like chemotaxis \cite{K-S}, ecology \cite{Gilad}, social systems \cite{Epstein}, electric circuits \cite{BPS},  drift-diffusion in semiconductors \cite{Chen}, chemical reactions \cite{GLSS}, turbolent transport in plasmas \cite{Castillo}, granular material \cite{Aranson}, and cell division in tumor growth \cite{Sherratt} and have also been extensively numerically investigated \cite{GGJ,BB,GLSAN}.

The plan of the paper is the following. In Section~\ref{sec:1} we perform a linear stability analysis close to the coexistence equilibrium, showing that cross-diffusion is the necessary ingredient for pattern formation. 	In Section~\ref{sec:weakly}, through a weakly nonlinear analysis, we derive the cubic Stuart-Landau equation (in the subcritical regime, however, we have to push the analysis up to the fifth order to derive the quintic Stuart-Landau equation), which allow us to predict the shape and the amplitude of the pattern. 
In the subcritical region, where the linear analysis prescribes only Turing pattern with no temporal oscillations, we numerically detect the coexistence of Turing pattern with a limit cycle. These findings confirm the result that, even without interaction with either a Hopf or a wave instability, the Turing instability together with cross diffusion in a predator-prey model can give rise to spatiotemporally oscillating solutions (\cite{LJLW}). Oscillating Turing patterns of similar type, have been also found in \cite{Aragon}.
Moreover, varying $\gamma$, which corresponds to moving away from the supercritical region, the oscillating pattern undergoes a series of bifurcations leading to torus, period doubling and chaos.
We defer to a forthcoming paper the investigation of the transition to the chaotic dynamics of system \eqref{pp_adim}, whose relevance in the modeling of natural phenomena has been pointed out by many authors \cite{MPTML,BP}.
Finally, in Section~\ref{sec:traveling} we consider the case of spatially modulated patterns, when the size of the spatial domain is large and the pattern is sequentially formed invading the whole domain as a traveling wavefront. In this case, we derive the real Ginzburg-Landau amplitude equation whose solution gives, to a good approximation, the shape and the speed of the traveling front.



\section{Turing instability}
\label{sec:1}
In this section  we shall investigate the conditions on the system \eqref{pp_adim} for the onset of Turing instability.
The coexistence equilibrium for the kinetics is $
(u_0,v_0) \equiv (1,r-\gamma) 
$
which is biologically significant if and only if 
$
r-\gamma>0
$.
When it exists, the equilibrium is stable and can be an attractive node (for $0<r<\gamma+\gamma^2/4$) or an attractive spiral (for $r>\gamma+\gamma^2/4$). Notice that this kinetics does not exhibit any Hopf bifurcation.
%
The linearized system in the neighborhood of $(u_0,v_0)$ is:
\begin{equation}
\label{linearizzazione}
\text{\textbf{w}}_t =\Gamma \text{J}\text{\textbf{w}}+D \nabla^2\text{\textbf{w}}, \qquad \text{\textbf{w}}=
\begin{pmatrix}
u-u_0\\
v-v_0
\end{pmatrix},
\end{equation}
where
\begin{equation}
J=
\begin{pmatrix}
-\gamma && -1\\
r-\gamma && 0
\end{pmatrix}, \qquad 
D=
\begin{pmatrix}
1 && 0\\
d_{21}v_0 && d_2
\end{pmatrix}.
\end{equation}

\begin{figure*}
\center
\includegraphics[width=\textwidth,height=6.5cm]{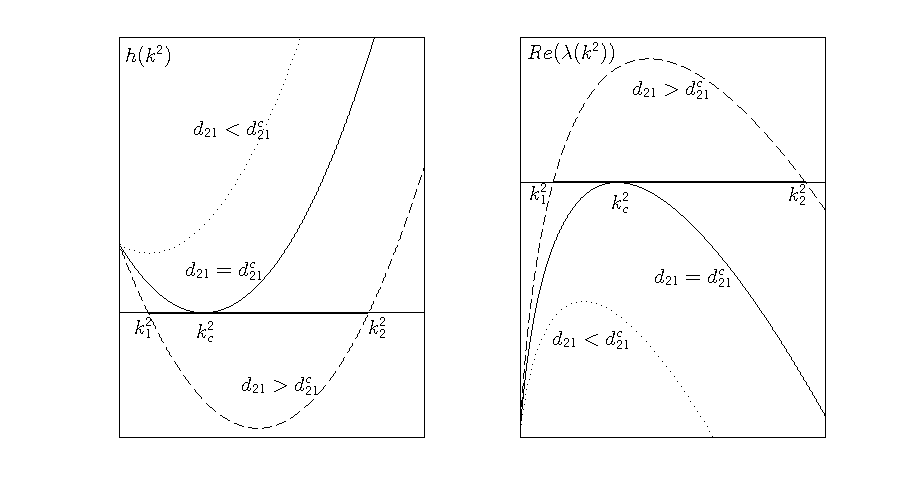}
\caption{\small{Left: Plot of $h(k^2)$ for different values of $d_{21}$. Right: Plot of the real part of the growth rate of the kth mode. A band of growing mode is present for $d_{21}>d_{21}^c$.}}
\label{fig:hrel}
\end{figure*}

The dispersion relation, which gives the eigenvalue $\lambda$ as a function of the wavenumber $k$, reads:
\begin{equation}
\label{PolCar}
\lambda^2+(-\Gamma \text{tr}(J)+k^2 \text{tr}(D))\lambda+ h(k^2)=0,
\end{equation}
where
\begin{equation}
\label{h}
h(k^2)=k^4\text{det}(D)+k^2\Gamma q+ \Gamma^2 \det(J),
\qquad\textrm{with} \quad
q=\gamma d_2 -\det(J)d_{21}.
\end{equation}
For the steady state to be linearly unstable to spatial disturbances we require $\text{Re}(\lambda(k))>0$ for some $k \neq 0$. Since $(u_0,v_0)$ is stable for the kinetics, one has that $\text{tr}(J)<0$. Moreover $\text{tr}(D)>0$. Therefore we are looking for those modes $k$ for which $h(k^2)<0$.  The only possibility for $h(k^2)<0$ is $q<0$, which means:
\begin{equation}
\label{CN}
d_{21}>\frac{\gamma d_2}{\det(J)}.
\end{equation}
Thus, the only potential destabilizing mechanism is the presence of the cross-diffusion term, while predator linear diffusion plays a stabilizing role.
The condition for the marginal stability at some $k=k_c$ is
\begin{equation}
\label{min}
\min(h(k_c^2))=0
\end{equation} 
and the minimum of $h$ is attained when
$
k^2_{min}=-\Gamma q/(2 \det(D)).
$
As shown in Figure \ref{fig:hrel}, the graph of $h(k^2)$ depends on $d_{21}$, which plays the role of the bifurcation parameter. Bifurcation happens at the critical value 
\begin{equation}
\label{d21c}
d_{21}^c=\frac{\gamma d_2+2\sqrt{\det(J)\det(D)}}{\det(J)},
\end{equation} 
in correspondence of the critical wavenumber
\begin{equation}
k_c^2=\Gamma\sqrt{\frac{\det(J)}{d_2}} \neq 0.
\end{equation}

For $d_{21}>d_{21}^c$ the system admits a finite $k$ pattern-forming stationary instability. The unstable wavenumbers stay in between the roots of $h(k^2)$, denoted by $k_1^2$ and $k_2^2$, which are proportional to $\Gamma$. Hence, to guarantee the emergence of spatial pattern, $\Gamma$ must be big enough so that at least one of the modes allowed by the boundary conditions, falls within the interval $[k_1^2,k_2^2]$.
Thus, the conditions under which patterns arise are:
\begin{equation}
r-\gamma>0, \qquad d_{21}>d_{21}^c,
\end{equation}
with $d_{21}^c$ given by \eqref{d21c}. Since the growth rate $\lambda$ crosses the zero having its imaginary part equal to zero ($\lambda(k_c)=0$), linear analysis predicts a stationary (Turing) bifurcation. Moreover, since $Im(\lambda)\neq0$ only for $0\le k<k_1$, from the results of the linear analysis we expect stationary patterns. 

\section{Weakly nonlinear analysis}
\label{sec:weakly}

 In this Section we shall perform a weakly nonlinear analysis based on the method of multiple scales.
Let $\varepsilon^2=(d_{21}-d_{21}^c)/d_{21}^c$ be the dimensionless distance from the threshold. 
For notational convenience, we shall denote with $d$  the bifurcation parameter $d_{21}$.

 We introduce new scaled coordinates, which will be treated as distinct variables, in addition to the original ones: 
%
\begin{equation}
\label{slow_var}
T=\varepsilon^2t, \qquad X=\varepsilon x
\end{equation}
and therefore the space and time derivatives decouple as $\partial_t\to \partial_t+ \varepsilon^2 \partial_T$ and $\partial_x\to \partial_x+ \varepsilon \partial_X$.
At this stage we shall not consider the possibility of spatial slow modulation. This will be done in Section~\ref{sec:traveling}.

Separating the linear part from the nonlinear one, we can recast the system \eqref{pp_adim} for the perturbation $\text{\textbf{w}}=(\text{w}^u,\text{w}^v)$, in the following form:
\begin{equation}
\label{separ}
\partial_t \text{\textbf{w}} = \mathcal{L}^{d} \text{\textbf{w}} + \frac{1}{2}\mathcal{Q}_K (\text{\textbf{w}},\text{\textbf{w}})+d
\begin{pmatrix}
0\\
\text{w}^v \nabla^2 \text{w}^u+\nabla \text{w}^v \cdot  \nabla \text{w}^u
\end{pmatrix},
\end{equation}
where $\mathcal{L}^d=\Gamma J+D^d \nabla^2$ is a linear operator depending on the bifurcation parameter $d$ and the bilinear operator $\mathcal{Q}_K$, which represents the nonlinear parts of the kinetics, is defined as: 
\begin{equation}
\mathcal{Q}_K (\text{\textbf{x}},\text{\textbf{y}})=\Gamma
\begin{pmatrix}
-2\gamma \text{x}^u\text{y}^u-(\text{x}^u\text{y}^v+\text{x}^v\text{y}^u)\\
\text{x}^u\text{y}^v+\text{x}^v\text{y}^u
\end{pmatrix},
\end{equation}
\noindent  where $\text{\textbf{x}}=(\text{x}^u,\text{x}^v)$ and $\text{\textbf{y}}=(\text{y}^u,\text{y}^v)$. Finally, the last term in \eqref{separ} is the nonlinear diffusion term.

 We expand $d$ and $\text{\textbf{w}}$ as:
\begin{eqnarray}
d&=&d^c+\varepsilon^2d^{(2)}+O(\varepsilon^4), \label{espd}\\
\text{\textbf{w}}&=&\varepsilon\text{\textbf{w}}_1+\varepsilon^2\text{\textbf{w}}_2+\varepsilon^3\text{\textbf{w}}_3+O(\varepsilon^4) \label{espw}.
\end{eqnarray}
Accordingly, the operators $\mathcal{L}^d$ and $\mathcal{Q}_K$ can be expanded as:
\begin{eqnarray}
\mathcal{L}^d&=&\mathcal{L}^{d^c}+ \varepsilon^2 d^{(2)}
\begin{pmatrix}
0 && 0\\
v_0 && 0
\end{pmatrix} \nabla^2+O(\varepsilon^4),\\
\mathcal{Q}_K(\text{\textbf{w}},\text{\textbf{w}})&=&\varepsilon^2\mathcal{Q}_K(\text{\textbf{w}}_1,\text{\textbf{w}}_1)+2\varepsilon^3\mathcal{Q}_K(\text{\textbf{w}}_1,\text{\textbf{w}}_2)+O(\varepsilon^4),
\end{eqnarray}

Substituting all the above expansions into \eqref{separ} and collecting the terms at each order in $\varepsilon$, one gets a sequence of equations for the $\text{\textbf{w}}_i$.


At $O(\varepsilon)$ we recover the linear problem $\mathcal{L}^{d^c}\text{\textbf{w}}_1=0$, whose solution satisfying the Neumann boundary conditions is:
\begin{equation}
\text{\textbf{w}}_1=A(T)\boldsymbol{\rho}\cos(k_cx), \quad \text{with } \rho \in Ker(\Gamma J-k_c^2 D^{d^c}),
\end{equation}
where $A(T)$ is the amplitude of the pattern which is still arbitrary at this level, since $\mathcal{L}^{d^c}$ does not act on the slow scale $T$. The vector $\boldsymbol{\rho}=(\rho^u,\rho^v)$ is defined up to a constant and can be normalized in the following way:
\begin{equation}
\label{rho}
\boldsymbol{\rho}=
\begin{pmatrix}
1 \\ M
\end{pmatrix}, \quad \text{with } M=\frac{\Gamma J_{21}-k^2_c D_{21}^{d^c}}{-\Gamma J_{22}+k^2_c D_{22}^{d^c}},
\end{equation}
where $J_{ij}$ and $D_{ij}^{d^c}$ ($i,j=1,2$) are the $i,j$-entries of the matrices $J$ and $D^{d^c}$.

At $O(\varepsilon^2)$  we obtain the following system:
\begin{equation}
\label{F}
\mathcal{L}^{d^c}\text{\textbf{w}}_2=
-\frac{1}{4}A^2 \mathcal{Q}_K(\boldsymbol{\rho},\boldsymbol{\rho})+\biggl(-\frac{1}{4}\mathcal{Q}_K(\boldsymbol{\rho},\boldsymbol{\rho})
+d^c k_c^2 
\begin{pmatrix}
0 \\ M
\end{pmatrix}\biggl)\cos(2k_cx) .
\end{equation}

By the Fredholm alternative, eq. \eqref{F} admits solution if and only if $\Braket{\text{\textbf{F}},\boldsymbol{\tilde{\psi}}}=0$, where $\Braket{\cdot,\cdot}$ is the scalar product in $L^2(0,2\pi/k_c)$, and $\boldsymbol{\tilde{\psi}} \in Ker \{ (\mathcal{L}^{d^c})^\dagger \}$. Since
\begin{equation}
\label{psitilde}
\boldsymbol{\tilde{\psi}}=\boldsymbol{\psi} \cos(k_cx), \quad \text{with } \boldsymbol{\psi}=
\begin{pmatrix}
1 \\M^*
\end{pmatrix}\text{ and }M^*=\frac{\Gamma J_{12}-k^2_c D_{12}^{d^c}}{-\Gamma J_{22}+k^2_c D_{22}^{d^c}},
\end{equation}
Fredholm alternative is automatically satisfied. Thus, for the linearity of the problem, one gets the solution of \eqref{F} satisfying the Neumann boundary conditions:
\begin{equation}
\text{\textbf{w}}_2=A^2 \text{\textbf{w}}_{20}+A^2\text{\textbf{w}}_{22}\cos(2k_cx),
\end{equation}
where $\text{\textbf{w}}_{2i}$, $i=0,2$, are solutions of the following linear systems:
\begin{eqnarray}
\Gamma J(\text{\textbf{w}}_{20})&=&-\frac{1}{4}\mathcal{Q}_K(\boldsymbol{\rho},\boldsymbol{\rho}),\label{20}\\
(\Gamma J-4k_c^2D^{d^c})(\text{\textbf{w}}_{22})&=&-\frac{1}{4}\mathcal{Q}_K(\boldsymbol{\rho},\boldsymbol{\rho})+d^c k_c^2 \label{w22} 
\begin{pmatrix}
0 \\ M
\end{pmatrix}.
\end{eqnarray}


At $O(\varepsilon^3)$ one gets:
\begin{equation}
\label{G}
\mathcal{L}^{d^c}\text{\textbf{w}}_3=
\biggl( \frac{dA}{dT} \boldsymbol{\rho}+ A \text{\textbf{G}}^{(1)}_1 +A^3 \text{\textbf{G}}^{(3)}_1 \biggl)\cos(k_cx)+ A^3\text{\textbf{G}}_3, \cos(3k_cx)
\end{equation}
where
\begin{eqnarray}
&\text{\textbf{G}}^{(1)}_1&=d^{(2)}k_c^2 \begin{pmatrix} 0\\ v_0 \end{pmatrix}, \notag\\
&\text{\textbf{G}}^{(3)}_1&=-\mathcal{Q}_K(\boldsymbol{\rho},\text{\textbf{w}}_{20})-\frac{1}{2} \mathcal{Q}_K(\boldsymbol{\rho},\text{\textbf{w}}_{22})
+d^ck_c^2 \begin{pmatrix}0 \\ \text{\textbf{w}}_{20}^uM+\text{\textbf{w}}_{22}^v-\frac{1}{2}\text{\textbf{w}}_{22}^v \end{pmatrix},\notag\\
&\text{\textbf{G}}_3&=-\frac{1}{2}\mathcal{Q}_K(\boldsymbol{\rho},\text{\textbf{w}}_{22})+d^ck_c^2 \begin{pmatrix}0 \\ 3\text{\textbf{w}}_{22}^uM+\frac{3}{2}\text{\textbf{w}}_{22}^v \end{pmatrix}.\notag
\end{eqnarray}
The solvability condition for \eqref{G} gives the Stuart-Landau equation for the amplitude $A(T)$:
\begin{equation}
\label{SL}
\frac{dA}{dT}= \sigma A- L A^3,
\end{equation}
where the coefficients $\sigma$ and $L$ are:
\begin{equation}
\label{sigmaL}
\sigma=-\frac{(\text{\textbf{G}}^{(1)}_1,\boldsymbol{\psi})}{(\boldsymbol{\rho},\boldsymbol{\psi})}, \qquad
L=\frac{(\text{\textbf{G}}^{(3)}_1,\boldsymbol{\psi})}{(\boldsymbol{\rho},\boldsymbol{\psi})}.
\end{equation}

In the region of the parameter space where the pattern can develop, provided that $d>d^c$, (see Fig. \ref{fig:Turing_space}.a)), it is straightforward to prove that the coefficient $\sigma$ is always positive. On the other  hand, the Landau constant $L$ can be positive or negative, depending on the value of the system parameters. Thus the dynamics of the Stuart-Landau equation can be divided into two qualitatively different cases: the supercritical case, when $L>0$, and the subcritical one, when $L<0$. As the explicit expression of $L$ as a function of all the parameters is quite involved, it will not be given here; in figure \ref{fig:Turing_space}.a) we show the  numerically computed curve across which $L$ changes its sign.

\subsection{The supercritical case}
If the coefficients $\sigma$ and $L$ in \eqref{SL} are both positive, the bifurcation is supercritical.
In this case the stable equilibrium of the Stuart-Landau equation is $A_{\infty}=\sqrt{\sigma/L}$ and the asymptotic in time behavior of the solution is given by:
\begin{equation}
\label{wnlsolution}
\text{\textbf{w}}=\varepsilon \boldsymbol{\rho} \sqrt{\frac{\sigma}{L}} \cos(k_cx)+\varepsilon^2 \frac{\sigma}{L}(\text{\textbf{w}}_{20}+\text{\textbf{w}}_{22}\cos(2k_cx))+O(\varepsilon^3),
\end{equation} 
where $\boldsymbol{\rho}$ is defined by \eqref{rho} and $\text{\textbf{w}}_{2i}$, $i=0,2$, are the solutions of the  systems \eqref{20}-\eqref{w22}.
For the above solution to satisfy the Neumann boundary conditions, $k_c$ should be an integer or semi-integer. We therefore define $\bar{k}_c$ as the first integer or semi-integer to become unstable as $d$ crosses its critical value $d_c$, and take as the weakly nonlinear approssimation equation \eqref{wnlsolution} in which $k_c$ is replaced by $\bar{k}_c$.
\begin{figure*}
\includegraphics[width=\textwidth,height=5cm]{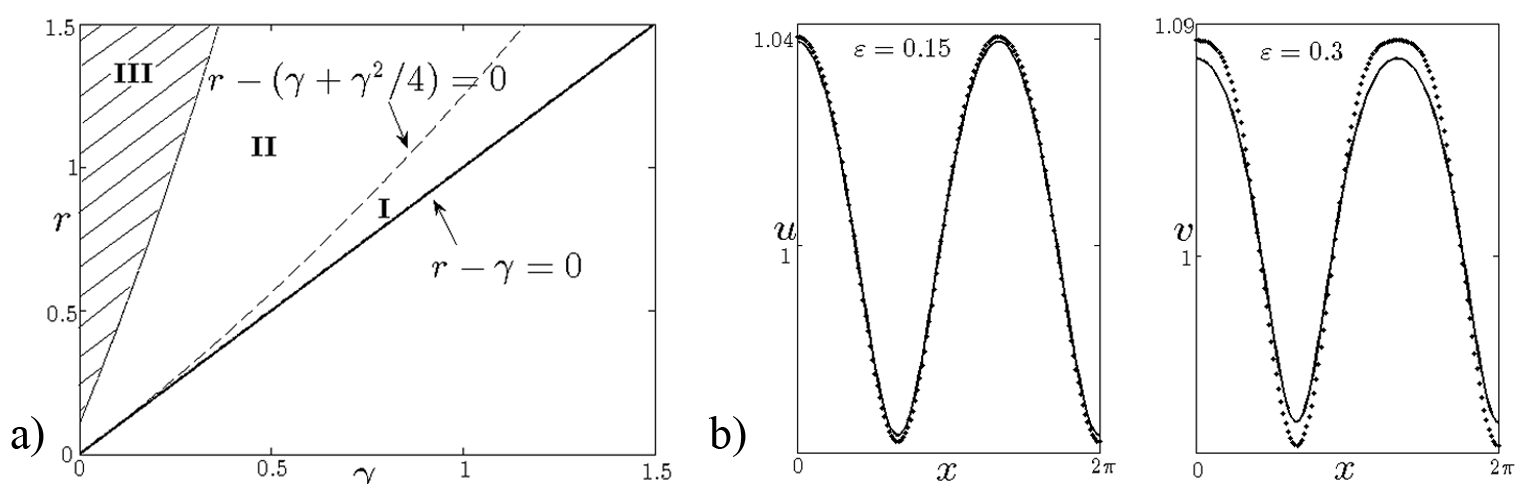}
\caption{\small{a) The Turing instability region, i.e. $r>\gamma$: the zones of supercritical bifurcation I (nodes) and II (spirals) and the subcritical bifurcation region III (spirals for the kinetics) are shown.  Across the curve that separates II from III, $L$ changes its sign. b) Comparison between the weakly nonlinear solution (dotted line) and the numerical solution of the system \eqref{pp_adim} (solid line) for $u$ (the plots of $v$ are similar). Parameters are $r=1.2$, $\gamma=1$, $d_2=1$, $\Gamma=5$, that result in $k_c\sim1.4953$ and $\bar{k}_c=1.5$. Left: $\varepsilon=0.15$ ($\varepsilon^2=0.0225$). Right: $\varepsilon=0.3$ ($\varepsilon^2=0.09$).}}
\label{fig:Turing_space}
\end{figure*}

In figure \ref{fig:Turing_space}.b) we show the comparison, for two different values of $\varepsilon$, between the stationary state predicted by the weakly nonlinear analysis (dotted line) and the stationary state reached from a random perturbation of the homogeneous equilibrium $(u_0,v_0)$, computed by solving numerically the system \eqref{pp_adim} (solid line). Notice that in the weakly nonlinear solution, we have choosen $d^{(2)}=d^c$, so that we measure the deviation from the critical value in relation to $d^c$.
Numerical results are in perfect agreement with what the weakly nonlinear analysis predicts. Infact, as the error is $O(\varepsilon^3)$, the distance evaluated in $L^2$ norm between the weakly nonlinear approssimation and the numerical solution, becomes an eighth as $\varepsilon$ is halved.  On the left, in which we have choosen $\varepsilon=0.15$, the distance is about $3.78 \times 10^{-3}$, on the rigth, for a larger deviation from the bifurcation value, the distance is about $2.06 \times10^{-2}$  and $\varepsilon=0.3$.

Thus in this case ($\sigma,L>0$) weakly nonlinear analysis is able to predict the stationary nature of the instability which leads to pattern 
 and the form and the amplitude of the pattern: we obtain stationary patterns both in spiral and node case. 

\subsection{The subcritical case}
When $L$ is negative, the bifurcation is subcritical:
  in this case the weakly nonlinear expansion has to be pushed up to the fifth order. 
 We therefore introduce the multiple time scales $T$ and $T_1$ as follows:
\begin{equation}
t=\frac{T}{\varepsilon^2}+\frac{T_1}{\varepsilon^4}+\dots,
\end{equation}
from which the time derivative decouples as $\partial_t \to \partial_t+\varepsilon^2 \partial_T+\varepsilon^4 \partial_{T_1}$, and expand $d$ and $\text{\textbf{w}}$ up to  the fifth order in $\varepsilon$.

Substituting these expansions into \eqref{separ}, up to $O(\varepsilon^3)$ we obtain the same equations presented in Section~\ref{sec:weakly}. At $O(\varepsilon^3)$, solvability condition $\Braket{\text{\textbf{G}},\boldsymbol{\tilde{\psi}}}=0$ for \eqref{G} leads again to \eqref{SL} for the amplitude, although now the derivative with respect to $T$ is a partial derivative, being $A=A(T,T_1)$. If it is satisfied, the solution is:
\begin{equation}
\text{\textbf{w}}_3=(A\text{\textbf{w}}_{31}+A^3\text{\textbf{w}}_{32})\cos(k_cx)+A^3\text{\textbf{w}}_{33}\cos(3k_cx),
\end{equation}
where $\text{\textbf{w}}_{3i}$, $i=1,2,3$, are solutions of the following linear systems:
\begin{eqnarray}
(\Gamma J-k_c^2D^{d^c})(\text{\textbf{w}}_{31})&=&\sigma\boldsymbol{\rho}-\text{\textbf{G}}^{(1)}_1, \notag\\
(\Gamma J-k_c^2D^{d^c})(\text{\textbf{w}}_{32})&=& -L\boldsymbol{\rho}+\text{\textbf{G}}^{(3)}_1,\notag\\
(\Gamma J-9k_c^2D^{d^c})(\text{\textbf{w}}_{33})&=&\text{\textbf{G}}_3.\notag
\end{eqnarray}

At $O(\varepsilon^4)$ the resulting equation is:
\begin{multline}
\mathcal{L}^{d^c}\text{\textbf{w}}_3=
2A\frac{\partial A}{\partial T}\text{\textbf{w}}_{20}+A^2 \text{\textbf{H}}_0^{(2)}+A^4 \text{\textbf{H}}_0^{(4)}+\\
+\biggl(2A\frac{\partial A}{\partial T}\text{\textbf{w}}_{22}+A^2 \text{\textbf{H}}_2^{(2)}+A^4 \text{\textbf{H}}_2^{(2)}\biggl)\cos(2k_cx)+A^4\text{\textbf{H}}_4\cos(4k_cx),\label{HH}
\end{multline}
where $\text{\textbf{H}}_0^{(2)}$, $\text{\textbf{H}}_0^{(4)}$, $\text{\textbf{H}}_2^{(2)}$, $\text{\textbf{H}}_2^{(4)}$, $\text{\textbf{H}}_4$ are explicitly computed in terms of the system parameters. As the expressions are quite involved, for conciseness we shall not give them here.
The solvability condition for \eqref{HH} is automatically satisfied and the solution is:
\begin{equation}
\text{\textbf{w}}_4=A^2\text{\textbf{w}}_{40}+A^4\text{\textbf{w}}_{41}+(A^2\text{\textbf{w}}_{42}+A^4\text{\textbf{w}}_{43})\cos(2k_cx)+A^4\text{\textbf{w}}_{44}\cos(4k_cx),
\end{equation}
where the $\text{\textbf{w}}_{4i}$, $i=0,\dots,4$, are obtained by solving the following linear systems:
\begin{eqnarray}
(\Gamma J)(\text{\textbf{w}}_{40})&=&2\sigma\text{\textbf{w}}_{20}+\text{\textbf{H}}_0^{(2)}, \notag\\
(\Gamma J)(\text{\textbf{w}}_{41})&=&-2L\text{\textbf{w}}_{20}+\text{\textbf{H}}_0^{(4)}, \notag\\
(\Gamma J-4k_c^2D^{d^c})(\text{\textbf{w}}_{42})&=&2\sigma\text{\textbf{w}}_{22}+\text{\textbf{H}}_2^{(2)}, \notag\\
(\Gamma J-4k_c^2D^{d^c})(\text{\textbf{w}}_{43})&=&-2L\text{\textbf{w}}_{22}+\text{\textbf{H}}_2^{(4)}, \notag\\
(\Gamma J-16k_c^2D^{d^c})(\text{\textbf{w}}_{44})&=&\text{\textbf{H}}_4. \notag
\end{eqnarray}

At $O(\varepsilon^5)$ we finally obtain:
\begin{multline}
\label{L}
\mathcal{L}^{d^c}\text{\textbf{w}}_3=
\biggl(\frac{\partial A}{\partial T_1}\boldsymbol{\rho}+\frac{\partial A}{\partial T}\text{\textbf{w}}_{31}+3A^2\frac{\partial A}{\partial T}\text{\textbf{w}}_{32} +A\text{\textbf{I}}_1+A^3\text{\textbf{I}}_1^{(3)}+A^5\text{\textbf{I}}_1^{(5)}\biggl)\cos(k_cx)+\\
+\biggl(3A^2\frac{\partial A}{\partial T}\text{\textbf{w}}_{33}+A^3\text{\textbf{I}}_3^{(3)}+A^5\text{\textbf{I}}_3^{(5)}\biggl)\cos(3k_cx)+A^5\text{\textbf{I}}_5^{(5)}\cos(5k_cx),
\end{multline}
where $\text{\textbf{I}}_1$, $\text{\textbf{I}}_1^{(3)}$, $\text{\textbf{I}}_1^{(5)}$ $\text{\textbf{I}}_3^{(3)}$, $\text{\textbf{I}}_3^{(5)}$, $\text{\textbf{I}}_5^{(5)}$ are explicitly computed in terms of the system parameters.
The solvability condition for \eqref{L} is
\begin{equation}
\label{isso}
\frac{\partial A}{\partial T_1}= A \tilde{\sigma}-A^3\tilde{L}+A^5\tilde{Q},
\end{equation}
where the coefficients are
\begin{equation}
\begin{split}
\tilde{\sigma}&
=-\frac{(\sigma\text{\textbf{w}}_{31}+\text{\textbf{I}}_1,\boldsymbol{\psi})}{(\boldsymbol{\rho},\boldsymbol{\psi})},\\
\tilde{L}&
=\frac{(-L\text{\textbf{w}}_{31}+3\sigma\text{\textbf{w}}_{32}+\text{\textbf{I}}_1^{(3)},\boldsymbol{\psi})}{(\boldsymbol{\rho},\boldsymbol{\psi})},\\
\tilde{Q}&
=\frac{(3L\text{\textbf{w}}_{32}-\text{\textbf{I}}_1^{(5)},\boldsymbol{\psi})}{(\boldsymbol{\rho},\boldsymbol{\psi})}.
\end{split} 
\end{equation}
Adding up \eqref{SL} to \eqref{isso}, one gets the quintic Stuart-Landau equation
\begin{equation}
\label{SL5}
\frac{\partial A}{\partial T}=\bar{\sigma}A-\bar{L}A^3+\bar{Q}L^5
\end{equation}
where:
\begin{equation}
\bar{\sigma}=\sigma+\varepsilon^2\tilde{\sigma}, \qquad
\bar{L}= L+\varepsilon^2\tilde{L}, \qquad
\bar{Q}=\varepsilon^2\tilde{Q}.
\end{equation}

\begin{figure}
\begin{minipage}{0.51\textwidth}
\flushleft
{\includegraphics[width=\textwidth,height=5.5cm]{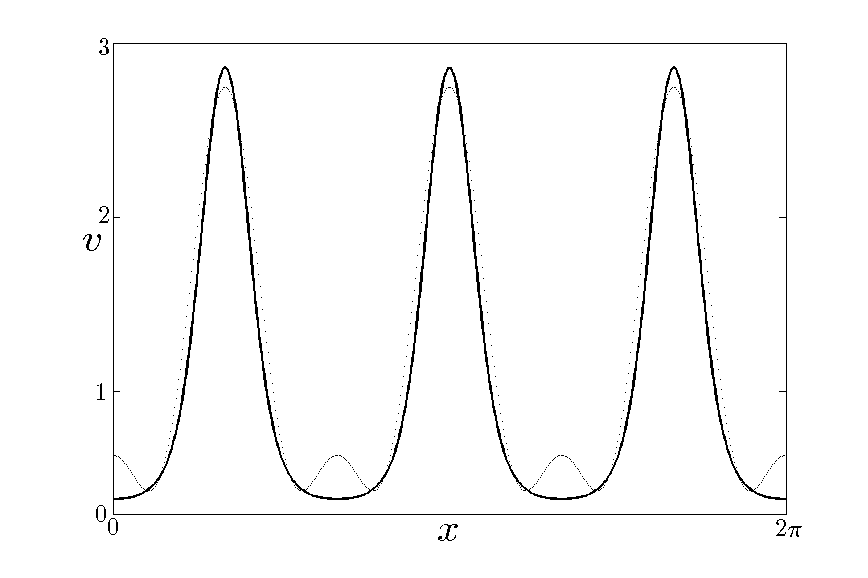}}
\end{minipage}
\begin{minipage}{0.5\textwidth}
\flushleft
{\includegraphics[width=.98\textwidth,height=5.5cm]{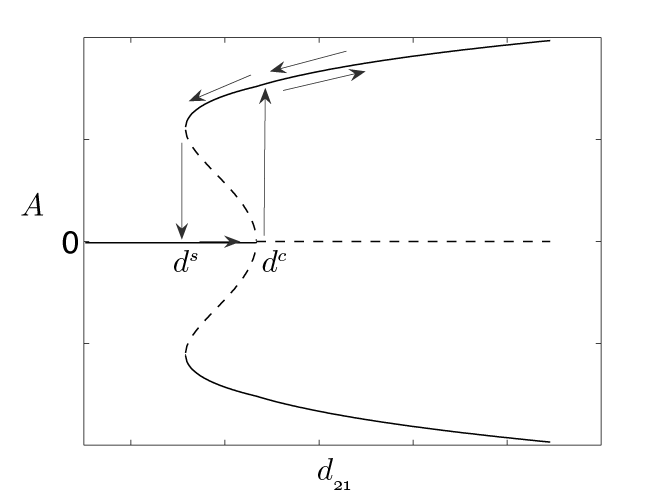}}
\end{minipage}
\caption{\small{Left: Comparison between the weakly nonlinear approximation (dotted line) and the numerical solution (solid line) of the sistem \eqref{pp_adim}. Parameters are $r=0.55$, $\gamma=0.1$, $d_2=3$, $\Gamma=10$, $\varepsilon^2=0.01$. Rigth: The corresponding bifurcation diagram.}}
\label{fig:quintico}
\end{figure}

In the subcritical case, namely when $\bar{\sigma}>0$ and $\bar{L}<0$, and when $\bar{Q}<0$, the study of the equilibria of \eqref{SL5} shows that there are two real stable equilibria, $A_{\infty,\pm}=\pm \sqrt{\frac{\bar{L}-\sqrt{\bar{L}^2-4\bar{\sigma}\bar{Q}}}{2\bar{Q}}}$, which represent the asymptotic values of the amplitude. We can distinguish three zones: for $d>d^c$ the only stable equilibria are $A_{\infty,\pm}$ and the origin is unstable.  For $d^s<d<d^c$ three different stable equilibra coexist: $A_{\infty,\pm}$ and the origin. Moreover there are two other equilibria which are unstable. Finally for $d<d^s$ the only stable equilibrium is the origin. The existence of different stable states, in corrispondence of a single value of $d$, allows for the possibility of hysteresis when varying $d$, as the numerical simulations  confirm. In Figure \ref{fig:quintico} we show a comparison between the weakly nonlinear approximation and the numerical solution of the sistem \eqref{pp_adim}. On 
the rigth we show the bifurcation diagram as a function of $d$.

\begin{figure*}
\includegraphics[width=\textwidth,height=5.5cm]{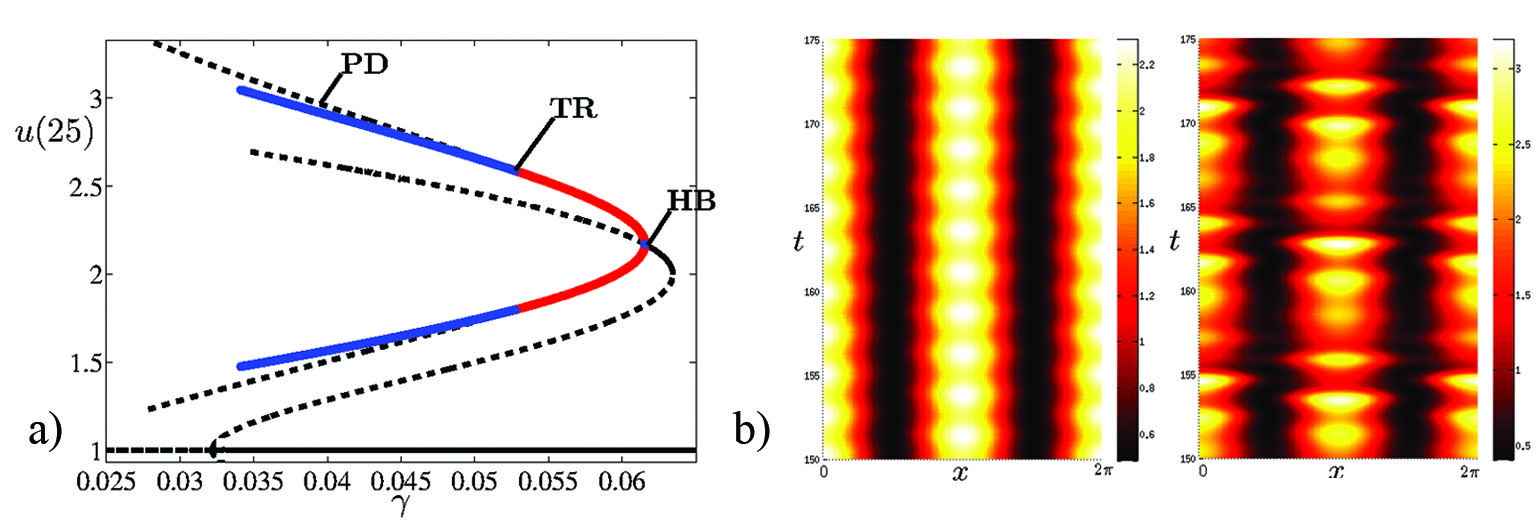}
\caption{\small{a) Bifurcation diagram of the central point $u(25)$ numerically calculated by AUTO with $r=0.85$, $\Gamma=10$, $d_2=1$, $d_{21}=2.269$. Successive bifurcations occur at $\gamma=0.03217$ (Subcritical Pitchfork), $\gamma=0.06153$ (Hopf Bifurcation),  $\gamma=0.05273$ (Torus Bifurcation) and $\gamma=0.03938$ (Period Doubling).
b) Time evolution of $u$ with $x\in [0,2\pi]$ on the horizontal axis and time increasing from bottom to top. The parameter set is as in a), with  $\gamma=0.0605$ on the left, and $\gamma=0.04391$ on the right. Only the final part of the trajectory is shown, after the effect of the transient have subsided.}}
\label{fig:ultimo}
\end{figure*}

However, while in the subcritical region which is close to the supercritical one the weakly nonlinear analysis is able to predict the form of the patterns which are still stationary, as $\gamma$ becomes smaller, going  deeper into the subcritical region, numerical simulations show the emergence of non-stationary oscillating pattern, whose basic form is still predicted by the weakly nonlinear analysis. As the parameter $\gamma$ is further decreased, the emergence of spatiotemporal chaos is recorded. To provide further insights into how the pattern changes as
the parameter $\gamma$ is varied, we have used the XPPAUT software package, solving the system \eqref{pp_adim} with a finite-difference scheme, and calculating the corresponding bifurcation diagram with the included sofware AUTO. Using a mesh of $50$ nodes, we get a set of $100$ ODEs solved through the stiff integrator CVODE. In Fig.\ref{fig:ultimo}.a) the numerically calculated bifurcation diagram of the central point of the species $u$, $u(25)$, is shown for the following values of the parameters: $r=0.85$, $\Gamma=10$, $d_2=1$, $d_{21}=2.269$.
The spatially uniform steady state loses its stability at $\gamma=0.03217$ through a subcritical pitchfork bifurcation. The bifurcating branch becomes stable at $\gamma=0.06343$ (Turing pattern) and undergoes a secondary Hopf bifurcation at $\gamma=0.06153$ (oscillatory Turing pattern). We remark the fact that, in this subset of the Turing space, no Hopf bifurcation is expected on the basis of the linear stability analysis. At smaller values of $\gamma$, the limit cycle loses stability and a torus bifurcation occurs at $\gamma=0.05273$. For smaller values of $\gamma$, chaotic oscillations are finally  observed (see Fig\eqref{fig:ultimo}.b) on the right). We believe that this bifurcation diagram is consistent with well-known scenarios of routes to chaos, whose details are the subject of current investigation and will be discussed in a forthcoming paper.

\section{Traveling fronts}
\label{sec:traveling}
When the size of the physical domain is large the pattern is formed sequentially and invades the entire domain as a travelling wavefront. 
To describe this phenomena one has to take into account the fast and the slow spatial dependence of the modulation, introducing in the weakly nonlinear analysis the slow spatial scale $X$ defined in \eqref{slow_var}.
Consequently one determines the expressions for the operators that appear in \eqref{separ}. At $O(\varepsilon)$ we recover the linear problem $L^{d^c}\text{\textbf{w}}_1=0$ where $L^{d^c}=\Gamma J+ D^d \partial_{xx}$ is the fast part of the linear operator $\mathcal{L}^{d}$. The solution is:
\begin{equation}
\label{solw1}
\text{\textbf{w}}_1=A(X,T)\boldsymbol{\rho}\cos(k_cx),
\end{equation}
when $\boldsymbol{\rho}$ is given in \eqref{rho} and $A$ is still arbitrary, as before.
\begin{figure}
\label{fig:wave}
\begin{center}
\includegraphics[width=\textwidth]{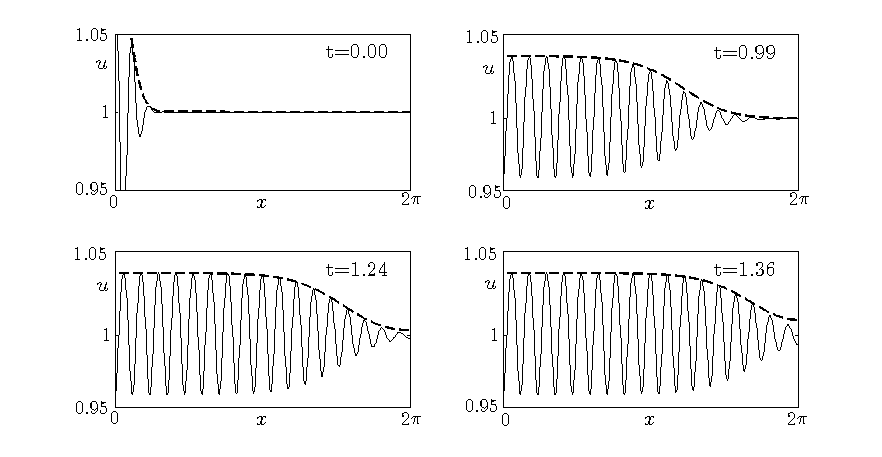}
\caption{\small{
The numerical solution of the Ginzburg-Landau equation \eqref{GL}  (dashed line) and the numerical solution of the system \eqref{pp_adim}, at different times. The parameters are all the same as in Figure \ref{fig:Turing_space}.b), with $\varepsilon=0.14$, except $\Gamma=650$.}}
\end{center}
\end{figure}

At $O(\varepsilon^2)$ the resulting equation is
\begin{multline}
\label{FF2}
L^{d^c}\text{\textbf{w}}_2=
-\frac{1}{4}A^2 \mathcal{Q}_K(\boldsymbol{\rho},\boldsymbol{\rho})
-\frac{1}{4}A^2\biggl(\mathcal{Q}_K(\boldsymbol{\rho},\boldsymbol{\rho})
-4d^c k_c^2 
\begin{pmatrix}
0 \\ M
\end{pmatrix}\biggl )\cos(2k_cx)+\\
+\biggl(2\frac{\partial A}{\partial X}k_c D^{d^c} \boldsymbol{\rho}\biggl) sen(k_cx)
\end{multline}
and the solvability condition is automatically satisfied. The solution is:
\begin{equation}
\text{\textbf{w}}_2=A^2 \text{\textbf{w}}_{20}+A^2\text{\textbf{w}}_{22}\cos(2k_cx)+\frac{\partial A}{\partial X} \text{\textbf{w}}_{X1} sen(k_cx),
\end{equation}
where $\text{\textbf{w}}_{2i}$, $i=0,2$, and $\text{\textbf{w}}_{X1}$ are solutions of the following linear systems:
\begin{eqnarray}
\Gamma J(\text{\textbf{w}}_{20})&=&-\frac{1}{4}\mathcal{Q}_K(\boldsymbol{\rho},\boldsymbol{\rho}),\notag \\
(\Gamma J-4k_c^2D^{d^c})(\text{\textbf{w}}_{22})&=&-\frac{1}{4}\mathcal{Q}_K(\boldsymbol{\rho},\boldsymbol{\rho})+d^c k_c^2 \label{22}
\begin{pmatrix}
0 \\ M
\end{pmatrix},\notag \\
(\Gamma J-k_c^2D^{d^c}) (\text{\textbf{w}}_{X1})&=&2 k_cD^{d^c} \boldsymbol{\rho} \label{X1}. \notag
\end{eqnarray}

At $O(\varepsilon^3)$ we find:
\begin{multline}
\label{GGG}
L^{d^c}\text{\textbf{w}}_3
= \biggl( \frac{\partial A}{\partial T} \boldsymbol{\rho}+\frac{\partial^2 A}{\partial X^2}\boldsymbol{\mathcal{G}}^{(XX)}_1 +A \boldsymbol{\mathcal{G}}^{(1)}_1 +A^3 \boldsymbol{\mathcal{G}}^{(3)}_1 \biggl) \cos(k_cx)+\\
 +A^3\boldsymbol{\mathcal{G}}_3 \cos(3k_cx)+\biggl( A \frac{\partial A}{\partial X} \boldsymbol{\mathcal{G}}^{(1X)}_2\biggl) sen(2k_cx)
\end{multline}
and
\begin{align}
&\boldsymbol{\mathcal{G}}^{(XX)}_1=-2k_cD^{d^c}\text{\textbf{w}}_{X1}-D^{d^c} \boldsymbol{\rho},\notag \\
&\boldsymbol{\mathcal{G}}^{(1)}_1=d^ck_c^2 \begin{pmatrix} 0\\ v_0 \end{pmatrix}, \notag\\
&\boldsymbol{\mathcal{G}}^{(3)}_1=-\mathcal{Q}_K(\boldsymbol{\rho},\text{\textbf{w}}_{20})-\frac{1}{2} \mathcal{Q}_K(\boldsymbol{\rho},\text{\textbf{w}}_{22})
+d^ck_c^2 \begin{pmatrix}0\\ \text{\textbf{w}}_{20}^uM+\text{\textbf{w}}_{22}^v-\frac{1}{2}\text{\textbf{w}}_{22}^v \end{pmatrix},\notag\\
&\boldsymbol{\mathcal{G}}_3=-\frac{1}{2}\mathcal{Q}_K(\boldsymbol{\rho},\text{\textbf{w}}_{22})+d^ck_c^2 \begin{pmatrix}0\\ 3\text{\textbf{w}}_{22}^uM+\frac{3}{2}\text{\textbf{w}}_{22}^v \end{pmatrix},\notag\\
&
\boldsymbol{\mathcal{G}}^{(1X)}_2=-\frac{1}{2}\mathcal{Q}_K(\boldsymbol{\rho},\text{\textbf{w}}_{X1})+d^c\begin{pmatrix}0 \\ 2k_cM+k_c^2(\text{\textbf{w}}_{X1}^uM+\text{\textbf{w}}_{X1}^v) \end{pmatrix}+8k_cD^{d^c}\text{\textbf{w}}_{22}.\notag
\end{align}
Solvability condition then leads to real Ginzburg-Landau equation for the amplitude $A(X,T)$:
\begin{equation}
\label{GL}
\frac{\partial A}{\partial T}=\nu \frac{\partial^2 A}{\partial X^2} +\sigma A- L A^3,
\end{equation}
where $\sigma$ and $L$ are given by \eqref{sigmaL}. The diffusion coefficient $\nu$ is given by:
\begin{equation}
\nu=\frac{(2k_cD^{d^c}\text{\textbf{w}}_{X1}+D^{d^c}\boldsymbol{\rho},\boldsymbol{\psi})}{(\boldsymbol{\rho},\boldsymbol{\psi})},
\end{equation}
where $\psi$ is given by \eqref{psitilde} and $(\cdot,\cdot)$ is the standard scalar product.

The Real Ginzburg-Landau equation is able to describe the invasion of the pattern of the entire domain, as its solution is the evelope of the pattern. 

In figure \ref{fig:wave} we show, at different times, the pattern that forms from a localized perturbation of the equilibrium.

\end{document}